\documentclass[10pt]{article}
\usepackage{fullpage}
\usepackage{amsmath}
\usepackage{amssymb}
\usepackage{amsfonts}
\usepackage{comment}
\usepackage{color}
\usepackage{cite}
\usepackage{subcaption}
\usepackage{graphicx}
\usepackage{enumitem}

\def\##1{\underline{#1}}
\def\=#1{\underline{\underline{#1}}}

\def\+
#1{\underline{\bf #1}}
\def\*#1{\underline{\underline{\bf #1}}}

\def\r#1{(\ref{#1})}
\def\l#1{\label{#1}}
\def\c#1{\cite{#1}}

\def\le{\left(}
\def\ri{\right)}
\def\les{\left[}
\def\ris{\right]}
\def\lec{\left\{}
\def\ric{\right\}}
\def\lek{[{\kern 0.1em}}
\def\rik{{\kern 0.1em}]}

\def\.{\mbox{ \tiny{$^\bullet$} }}

\def\eps{\varepsilon}

\def\epso{\eps_{\scriptscriptstyle 0}}
\def\lambdao{\lambda_{\scriptscriptstyle 0}}

\def\ko{k_{\scriptscriptstyle 0}}

\def\degC{\,^\circ{\rm C}}

\begin{document}

\begin{center}

\LARGE{ {\bf Thermal   hysteresis in amplification and attenuation of surface-plasmon-polariton waves
}}
\end{center}
\begin{center}
\vspace{10mm} \large
 
 {Tom G. Mackay}\footnote{E--mail: T.Mackay@ed.ac.uk.}\\
{\em School of Mathematics and
   Maxwell Institute for Mathematical Sciences\\
University of Edinburgh, Edinburgh EH9 3FD, UK}\\
and\\
 {\em NanoMM~---~Nanoengineered Metamaterials Group\\ Department of Engineering Science and Mechanics\\
Pennsylvania State University, University Park, PA 16802--6812,
USA}
 \vspace{3mm}\\
 Tran Vinh Son \\
\emph{ Department of Physics, Concordia University, Montreal H3G 1M8, Canada}
 \vspace{3mm}\\
 Alain Hach\'e \\
\emph{ D\'epartement de physique et d'astronomie, Universit\'e de Moncton, E1A 8T1, Canada} \vspace{3mm} \\
 {Akhlesh  Lakhtakia}\\
 {\em NanoM~---~Nanoengineered Metamaterials Group\\ Department of Engineering Science and Mechanics\\
Pennsylvania State University, University Park, PA 16802--6812, USA}

\normalsize

\end{center}

\begin{center}
\vspace{5mm} {\bf Abstract}
\end{center}

The propagation of surface-plasmon-polariton (SPP) waves at the planar interface
of a metal and a dielectric material
 was investigated for a  dielectric material with strongly temperature-dependent constitutive properties.  
 The metal was silver and the dielectric material was vanadium multioxide impregnated with a combination of active dyes. Depending upon the volume fraction of vanadium multioxide, either attenuation or amplification of  the SPP waves may be achieved; the degree of attenuation or amplification is strongly dependent on both the temperature and whether the temperature is increasing or decreasing.
 At intermediate volume fractions of vanadium multioxide, for a fixed temperature, 
  a SPP wave may experience attenuation if the temperature is increasing but experience amplification if the temperature is decreasing.

 \vspace{5mm}
\vspace{5mm}
 
\section{Introduction}

The planar interface of a plasmonic material and dielectric material guides the
propagation of surface-plasmon-polariton (SPP) waves \c{Boardman,Pitarke,ESW_book}. 
As the propagation of SPP waves is acutely sensitive to the constitutive properties of the plasmonic and dielectric materials involved, these surface waves are widely exploited in optical sensing applications \c{Homola}. The prospect of harnessing dielectric materials whose constitutive properties are strongly temperature dependent opens up possibilities of further applications for SPP waves in reconfigurable and multifunctional devices \c{Wang,Maguid,Huang,Waleed}.

At visible wavelengths, vanadium dioxide is a dissipative dielectric material whose constitutive properties are acutely sensitive to temperature  over the range  $25\degC$--$80\degC$ \c{Seo,Cueff,Lu,Shi,Cormier}.
Indeed, the crystal structure of  vanadium dioxide is monoclinic at temperatures below $58\degC$ and 
  tetragonal at temperatures above  $72\degC$ \c{Morin}, with
  both monoclinic and tetragonal crystals coexisting at intermediate  temperatures.
Furthermore, the temperature-induced monoclinic-to-tetragonal transition 
 is
hysteretic. 
The electromagnetic response of vanadium dioxide is  characterized
 by its  (complex-valued) relative
  permittivity $\eps_{\text{VO}}$, with $\mbox{Re} \lec \eps_{\text{VO}} \ric > 0 $ and $\mbox{Im} \lec \eps_{\text{VO}} \ric > 0 $ at visible wavelengths. The value 
  of $\eps_{\text{VO}}$ depends upon  temperature; also, 
  over the range $25\degC$--$80\degC$, it depends upon whether the material is being heated or cooled.
  {Parenthetically, the dissipative dielectric material-to-metal phase transition \c{Morin} that vanadium dioxide
  exhibits at free-space wavelength $\lambdao>1100$~nm \c{Kepic} is not relevant to our study.}

For optical  applications, thin films of vanadium dioxide  may often be desired \c{HW,STF_Book}. Such thin films are conveniently fabricated by a  vapor deposition process. However, depending upon the processing conditions and thickness of the film, the  deposition process may result in significant proportions of vanadium oxides other than vanadium dioxide being present in such films. Accordingly, in the absence of definitive stoichiometric evidence,  we shall refer these films as being composed of vanadium \emph{multioxide}.

Losses due to the dissipative nature of vanadium multioxide represent a potential impediment for optical applications. However, these losses may be overcome by mixing vanadium multioxide with an active material. Rhodamine dyes 
provide a class of suitable active materials that
are commonly used  to overcome losses at optical wavelengths in otherwise dissipative metamaterials \c{Sun,Campione}. {The use of active materials to amplify SPP waves is a well-established practice \c{Seidel,Torma,Liu,Berini}.}

Therefore, in the following, we investigate the temperature dependence  of SPP waves guided by the  interface of
(i) a homogenized mixture of 
vanadium multioxide and rhodamine dyes, and
 (ii) a plasmonic material which is taken to be silver. In particular, the thermal  hysteresis is  explored for both amplified and attenuated SPP waves.
The canonical boundary-value problem is considered in which 
SPP waves are guided by the interface $z=0$;
 the plasmonic material occupies the half-space $z<0$ and
the dielectric material occupies the half-space $z>0$.

\section{Relative permittivities of supporting materials}

\begin{figure}[!htb]
\centering
\includegraphics[width=16cm]{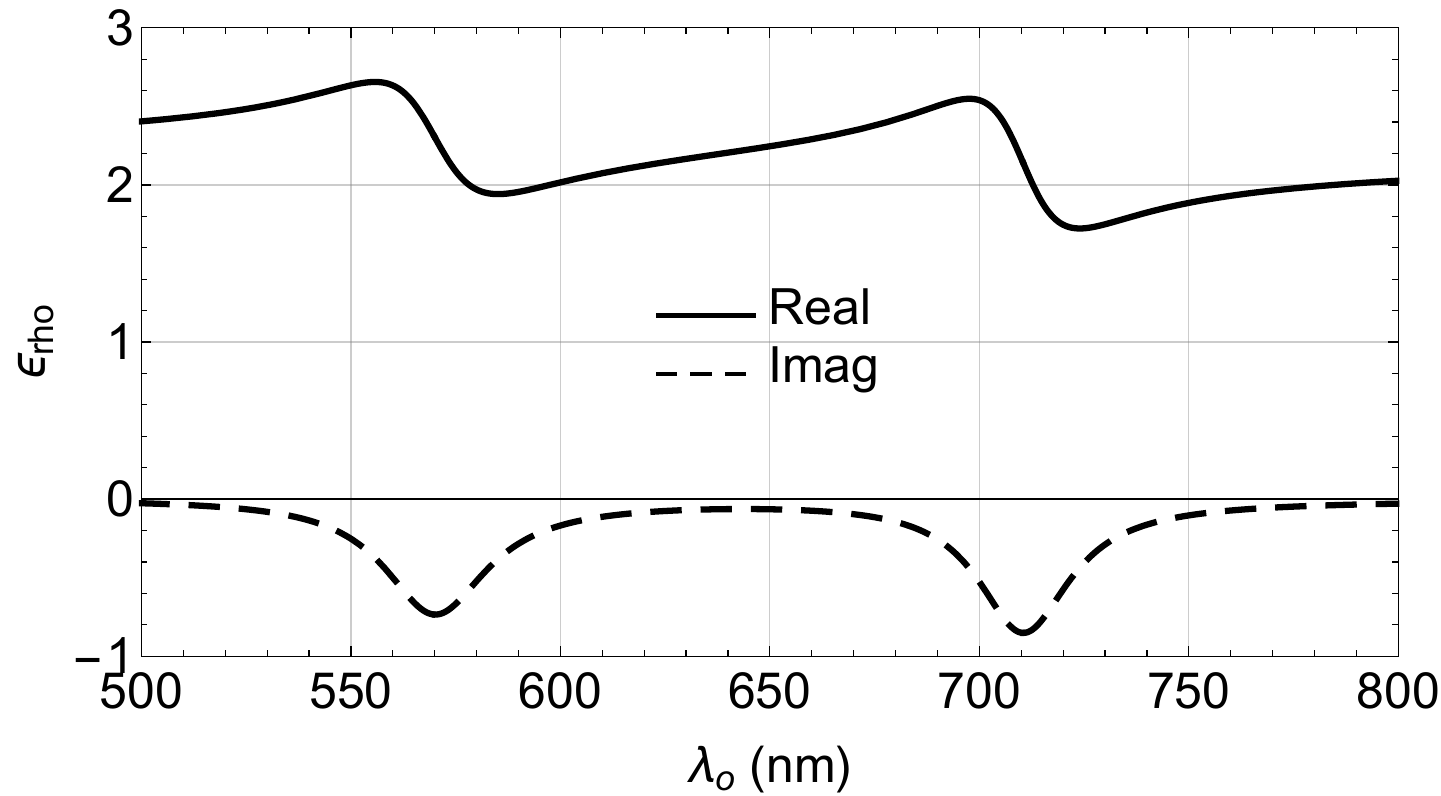}  
 \caption{\label{Fig1} Real and imaginary parts of $\eps_{\text{rho}}$ plotted against $\lambdao$, as determined from Eq.~\r{eps_rho}.
   }
\end{figure}

Following earlier works \c{Sun,Campione}, we consider a combination of 110 mM
rhodamine 6G and 25 mM rhodamine 800. The relative permittivity of the combination is estimated by the formula
\begin{equation} \l{eps_rho}
 \eps_{\text{rho}} =  \eps_r + \frac{1}{\epso} \le \frac{\sigma_{a1} \Delta N_1}{\omega^2 - \omega^2_{a1} + i \omega \Delta \omega_{a1}} 
+ \frac{\sigma_{a2} \Delta N_2}{\omega^2 - \omega^2_{a2} + i \omega \Delta \omega_{a2}} \ri,
\end{equation}
wherein the reference relative permittivity $\eps_r = 2.25$, the angular frequency is $\omega$, and the subscript labels `1' and `2' refer to rhodamine 6G and  rhodamine 800, respectively. The coupling strengths $\sigma_{a1} = 6.55 \times 10^{-8} \,\mbox{C}^2/\mbox{kg}$ and $ \sigma_{a2} = 1.71 \times 10^{-7} \,\mbox{C}^2/\mbox{kg} $; the center emission frequencies
$\omega_{a1} = 2 \pi f_{a1}$ with $f_{a1} = 526 \, \mbox{THz}$ and 
$\omega_{a2} = 2 \pi f_{a2}$ with$ f_{a2} = 422 \, \mbox{THz}$;
and the frequency linewidths $\Delta \omega_{a1} = 2 \pi \Delta \nu_{a1}$ with $\Delta \nu_{a1} = 27.7 \, \mbox{THz}$ and $\Delta \omega_{a2} = 2 \pi \Delta \nu_{a2}$ with $ \Delta \nu_{a2} = 15.9 \, \mbox{THz}$.
The factors
\begin{equation}
\Delta N_\ell = \frac{\le \tau^{(\ell)}_{21} - \tau^{(\ell)}_{10} \ri \Gamma^{(\ell)}_{\text{pump}}}{1+\le \tau^{(\ell)}_{32}+\tau^{(\ell)}_{21}+  \tau^{(\ell)}_{10} \ri \Gamma^{(\ell)}_{\text{pump}}} \bar{N}^{(\ell)}_0, \qquad \ell = 1, 2
\end{equation}
are determined by the energy level decay rates, the concentrations of 
 rhodamine 6G and  rhodamine 800, and the external pumping rates. Specifically, 
 the decay rates $\tau^{(1)}_{21} = 3.99 \, \mbox{ns}$, $ \tau^{(1)}_{10}  = \tau^{(1)}_{32} = 100 \, \mbox{fs}$, and $\tau^{(2)}_{21} = 500 \, \mbox{ps}, \quad \tau^{(2)}_{10}  = \tau^{(2)}_{32} = 100 \, \mbox{fs}$; the concentrations
 $\bar{N}^{(1)}_0 = 65 \times 10^{18} \, \mbox{cm}^{-3}$ and 
 $\bar{N}^{(2)}_0 = 15 \times 10^{18} \, \mbox{cm}^{-3}$; and the pumping rates $\Gamma^{(1)}_{\text{pump}} = 1.5 \times 10^{9} \, \mbox{s}^{-1}$ and 
 $\Gamma^{(2)}_{\text{pump}} = 6.5 \times 10^{9} \, \mbox{s}^{-1}$.
The real and imaginary parts of $ \eps_{\text{rho}} $ are plotted against   $\lambdao \in \les 500 \,\text{nm}, 800 \, \text{nm} \ris$ in Fig.~\ref{Fig1}. The minimum value of $\mbox{Im}\lec  \eps_{\text{rho}}  \ric$
arises at $\lambdao =  710$ nm. In order to maximally combat dissipative losses in vanadium multioxide,  we fix  the free-space wavelength at this value for the reminder of this study.
  
\begin{figure}[!htb]
\centering
\includegraphics[width=16cm]{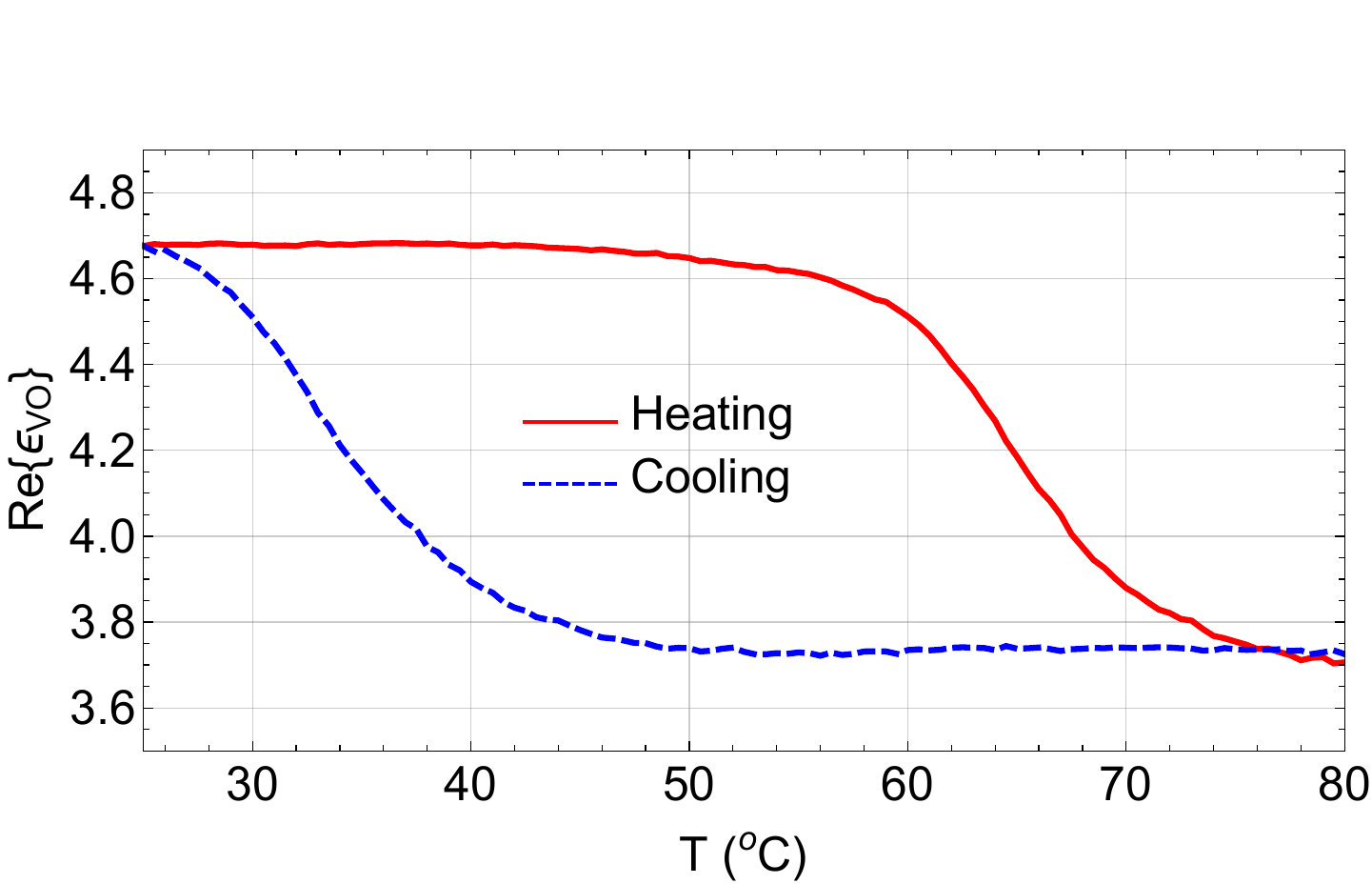}\\
\includegraphics[width=16cm]{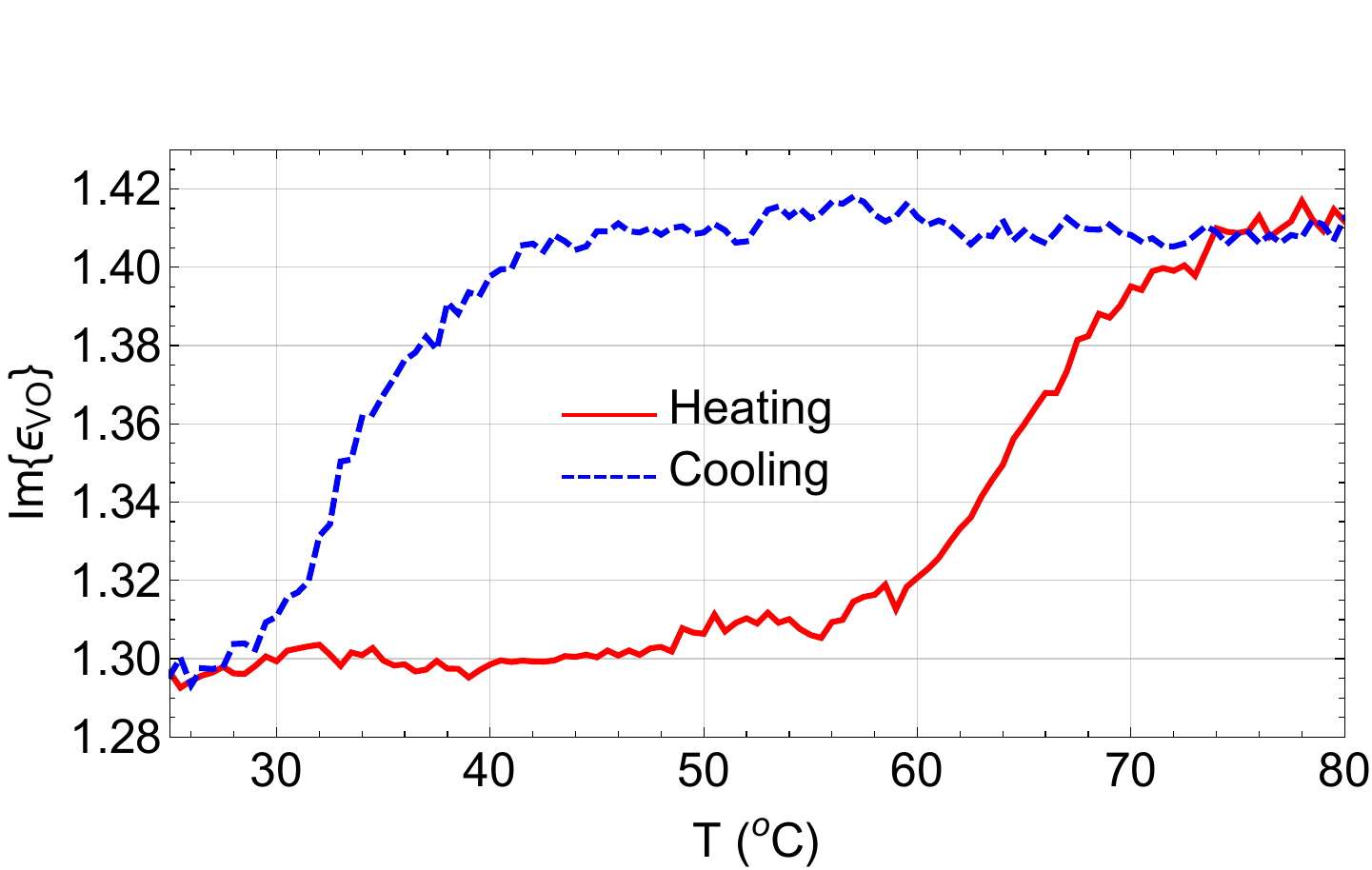}   
 \caption{\label{Fig2} Real and imaginary parts of $\eps_{\text{VO}}$ plotted against T for both heating and cooling phases, derived from  experimental values  \c{Tran}.  }
\end{figure}

  Plots of the real and imaginary parts  of $\eps_{\text{VO}}$ are provided in Fig.~\ref{Fig2} for the temperature $T$ range $\les 25\degC, 80\degC \ris$.
   These values were  derived by extrapolation of experimentally-determined values which were found at $\lambdao =800$ nm for both heating and cooling phases, following the method described in Ref.~\c{Tran}; and using 
  values determined by ellipsometry at $25^\circ$C and $95^\circ$C for $\lambdao = 710 $ nm.
  The hysteresis phenomenon
  displayed in Fig.~\ref{Fig2}
   extends over the range $25\degC < T < 75 \degC$ for both $\mbox{Re} \lec \eps_{\text{VO}} \ric$ and $\mbox{Im} \lec \eps_{\text{VO}} \ric$; over this range, $\mbox{Re} \lec \eps_{\text{VO}} \ric$ is larger for the heating phase than for the cooling phase whereas $\mbox{Im} \lec \eps_{\text{VO}} \ric$ is larger for the cooling phase than for the heating phase.
  The maximum difference in $\mbox{Re} \lec \eps_{\text{VO}} \ric$ between heating and cooling phases is approximately 0.9 and the
  maximum difference in $\mbox{Im} \lec \eps_{\text{VO}} \ric$ between heating and cooling phases is approximately 0.11.
  
  A homogenized mixture of vanadium multioxide, characterized by the relative permittivity
  $\eps_{\text{VO}}$ and volume fraction $f_{\text{VO}}$, 
   and a combination of  rhodamine dyes, characterized by the relative permittivity
  $\eps_{\text{rho}}$ and volume fraction $f_{\text{rho}}= 1  -
  f_{\text{VO}}$, occupies the half-space $z>0$.
  The relative permittivity of the homogenized  mixture, namely $\eps_{\text{mix}}$, is estimated using the Bruggeman homogenization formalism \c{Ward,MAEH}. 
  Accordingly, $\eps_{\text{mix}}$ is extracted from the Bruggeman equation
  \begin{equation} \l{eps_mix}
  f_{\text{rho}} \frac{\eps_{\text{rho}} - \eps_{\text{mix}}}{\eps_{\text{rho}}+2
  \eps_{\text{mix}}}
  + 
  f_{\text{VO}} \frac{\eps_{\text{VO}} - \eps_{\text{mix}}}{\eps_{\text{VO}}+2
  \eps_{\text{mix}}} = 0.
  \end{equation}
 {The electromagnetic response properties of vanadium multioxide is  assumed to be unchanged by the gain in the 
  rhodamine dyes, but the foregoing equation clearly shows the  gain to affect the electromagnetic response properties of the mixture of vanadium multioxide and rhodamine dyes. }
  
\begin{figure}[!htb]
\centering
\includegraphics[width=8cm]{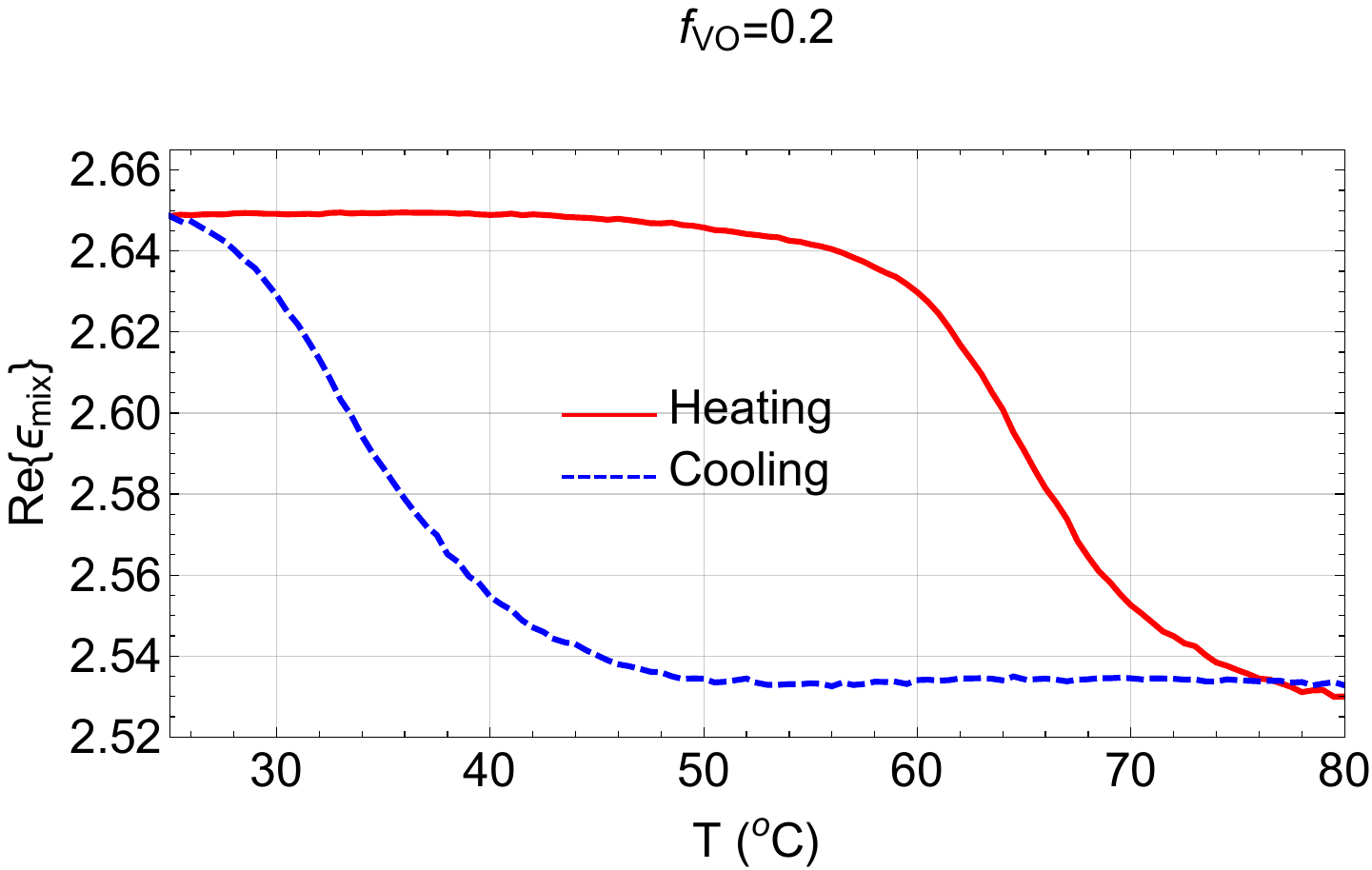}
\hfill
\includegraphics[width=8cm]{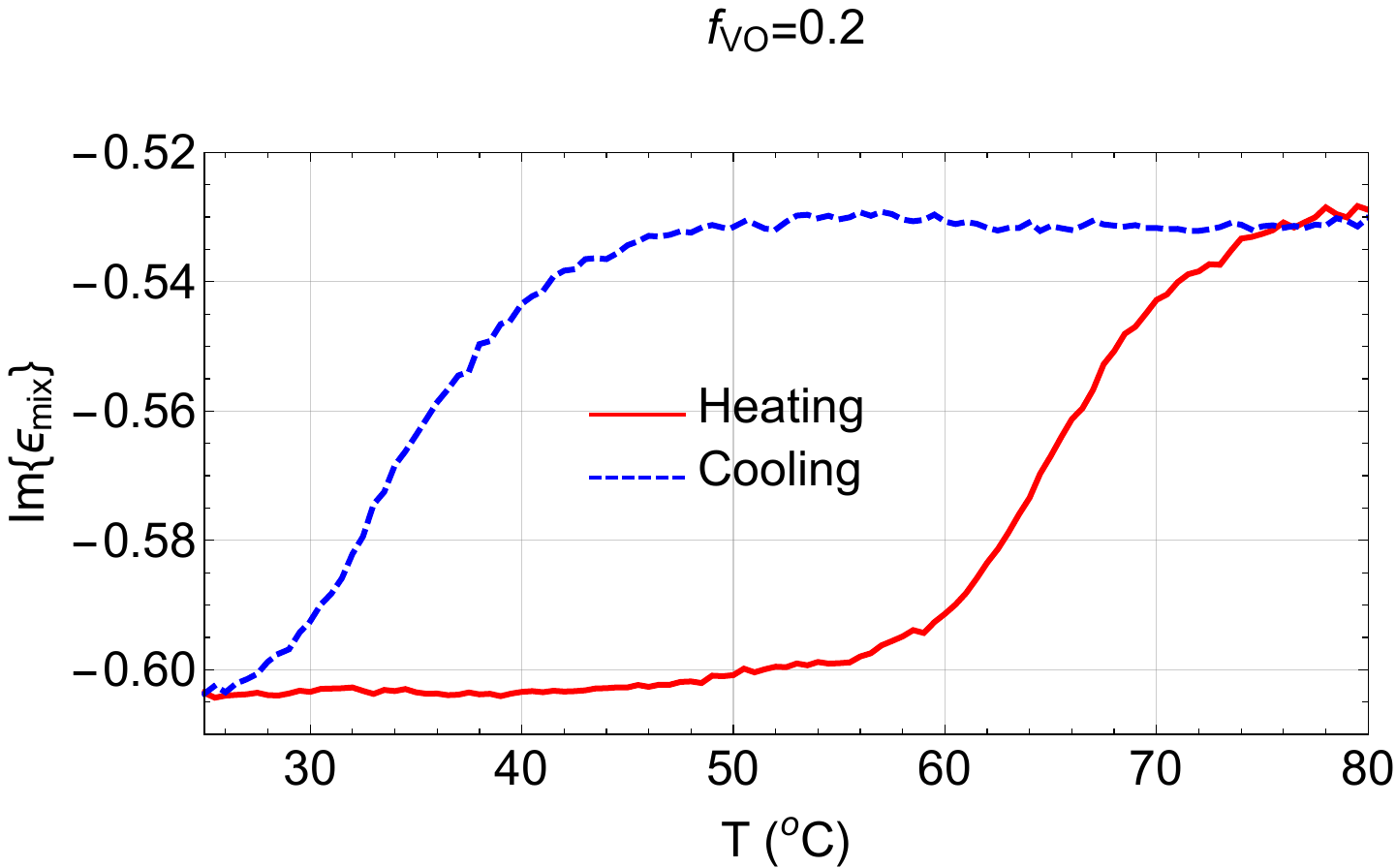}  \vspace{10mm}  \\
\includegraphics[width=8cm]{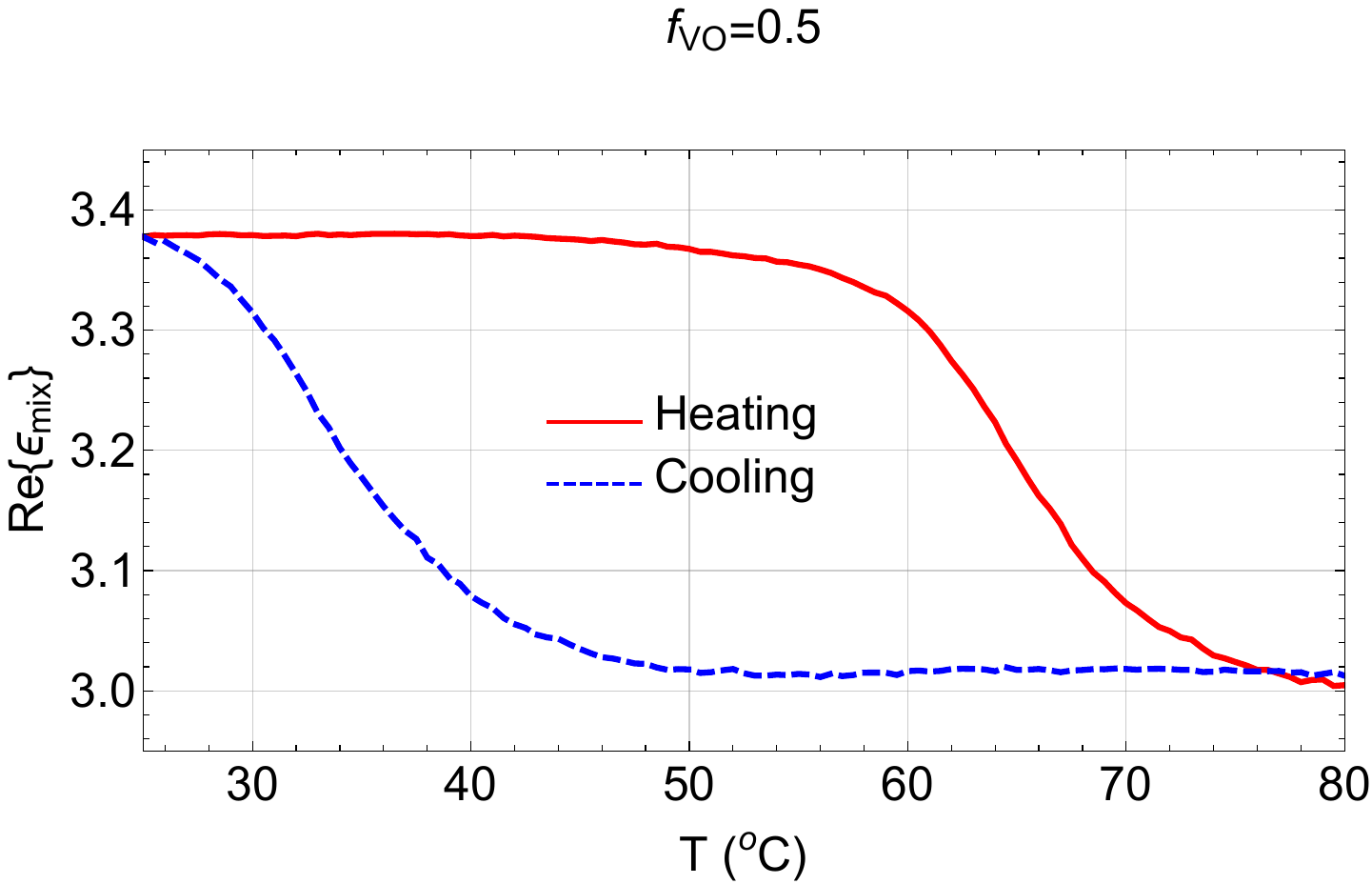}
\hfill
\includegraphics[width=8cm]{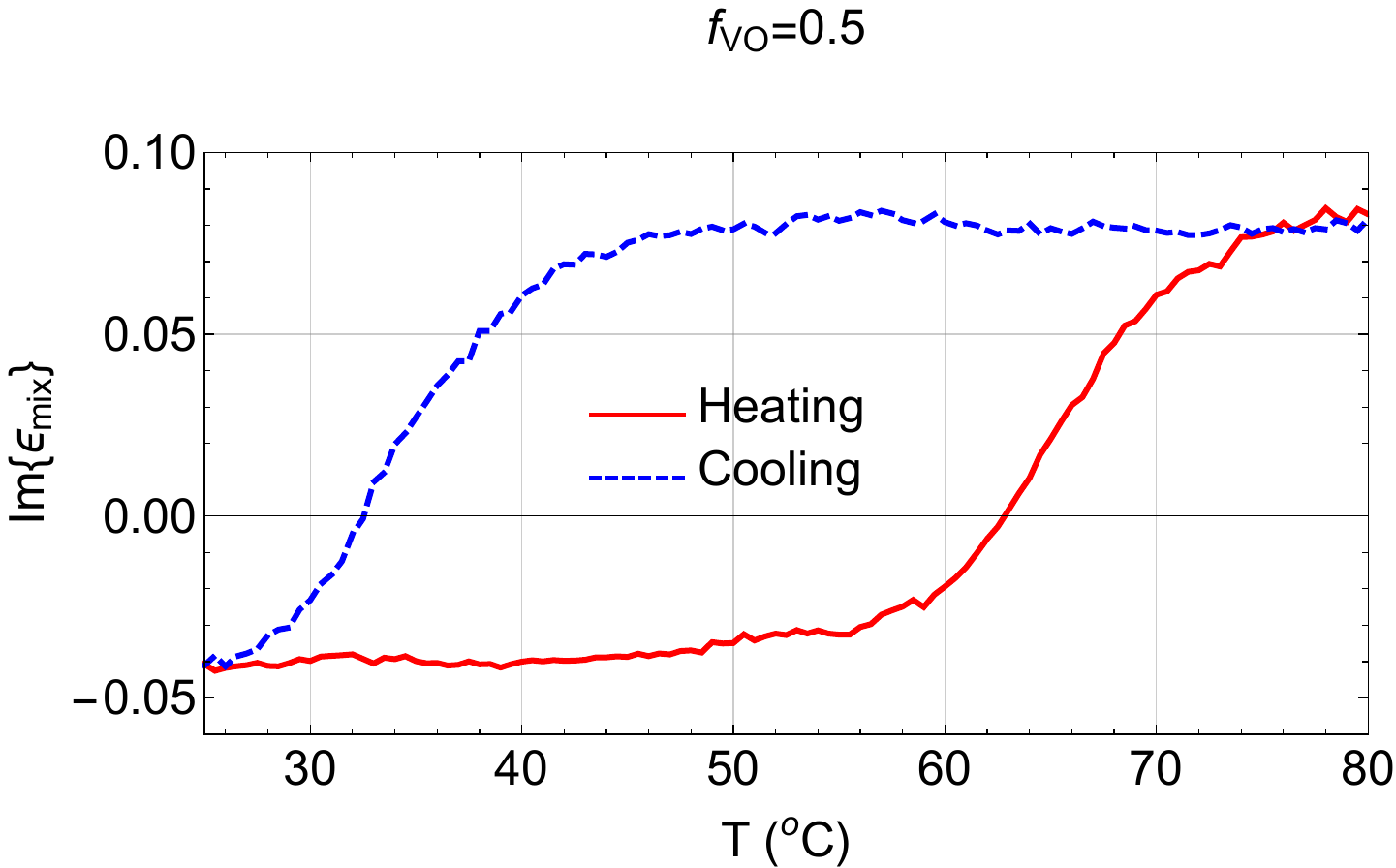} \vspace{10mm} \\
\includegraphics[width=8cm]{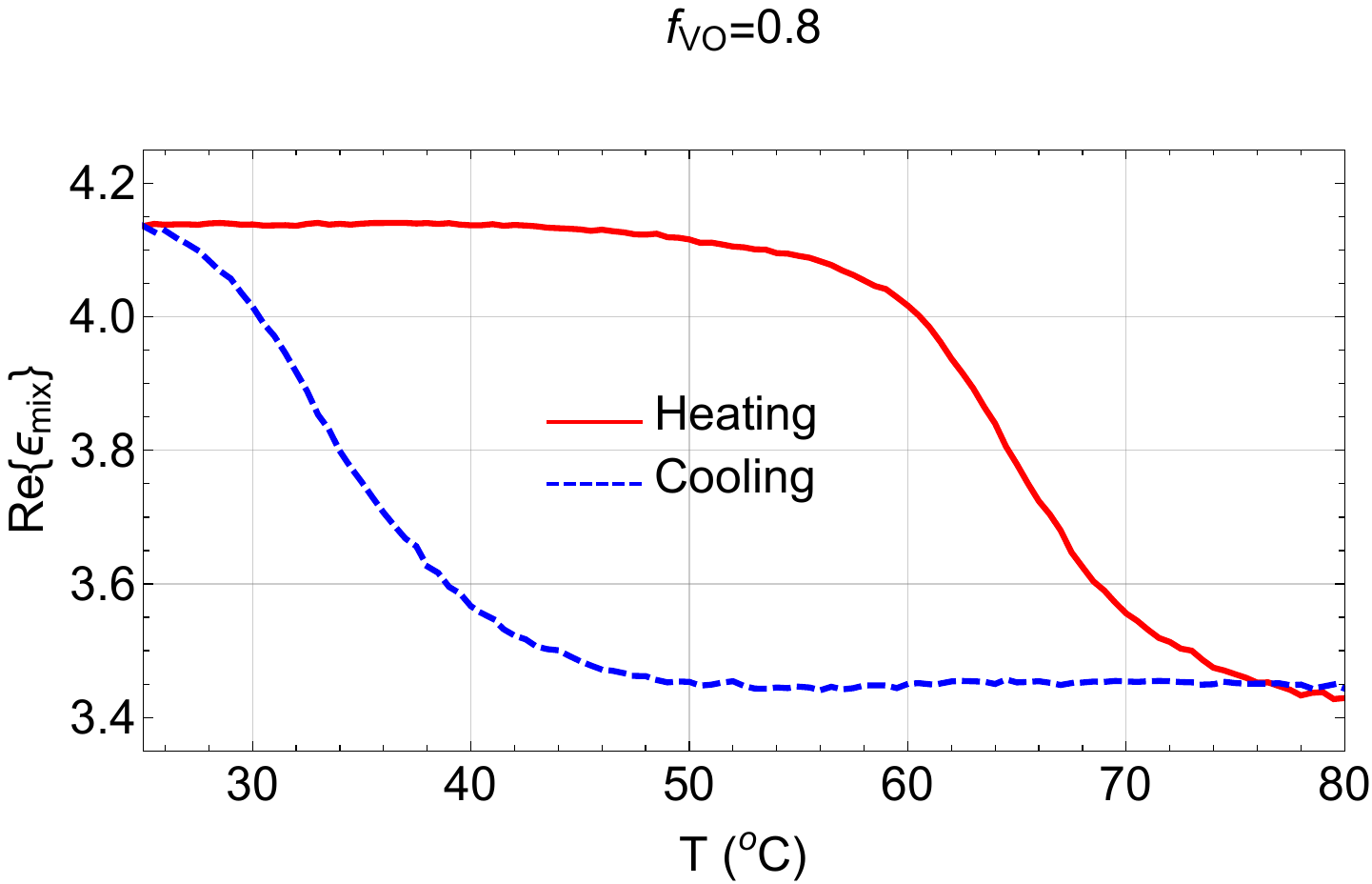}
\hfill
\includegraphics[width=8cm]{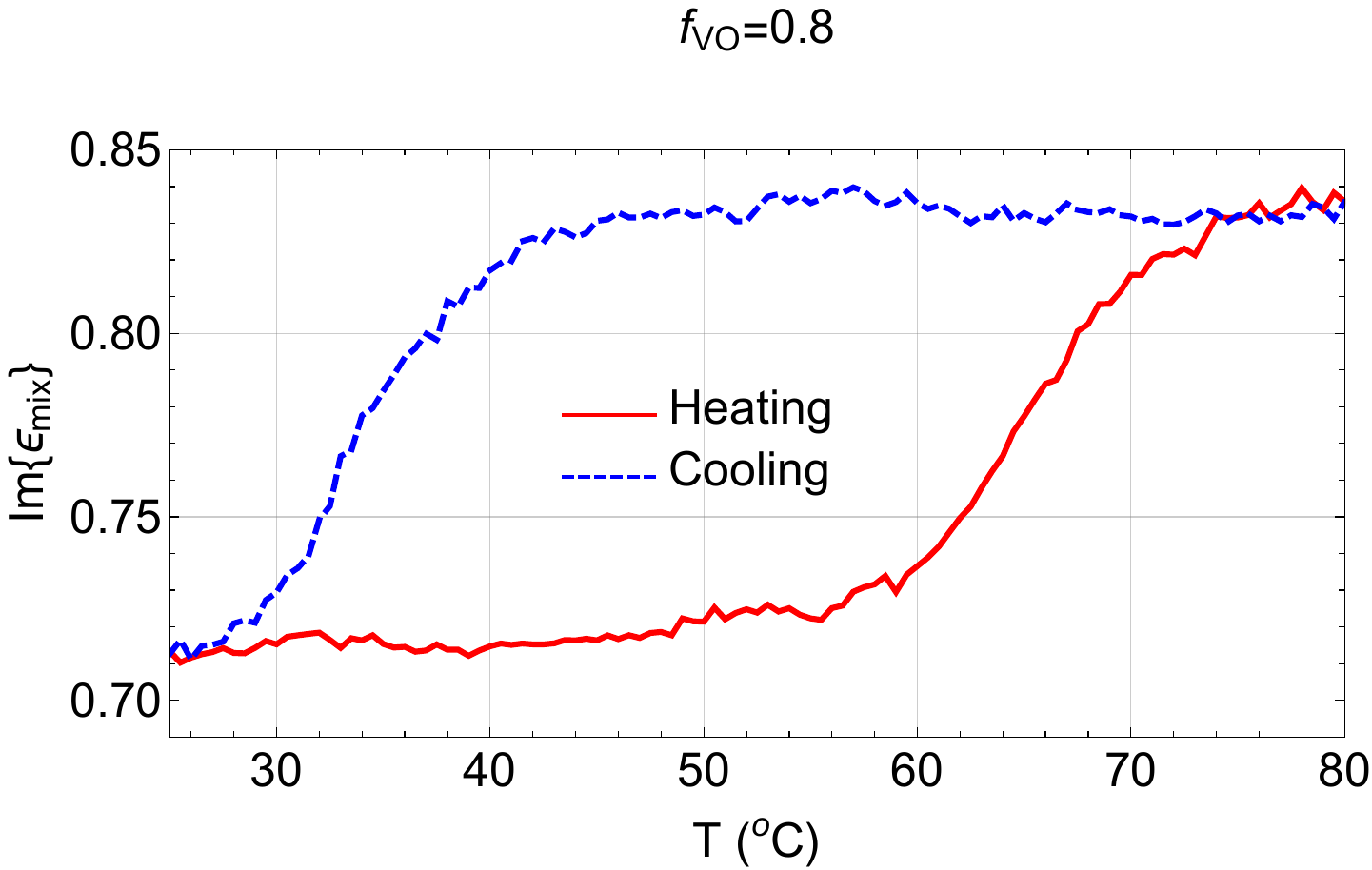}  
 \caption{\label{Fig3} 
 Real and imaginary parts of $\eps_{\text{mix}}$ plotted against $T$
 for both heating and cooling phases,
  as determined from the Bruggeman Eq.~\r{eps_mix} for $f_{\text{VO}}= 0.2$, 0.5, and 0.8.
   }
\end{figure}

  Plots of the real and imaginary parts of $\eps_{\text{mix}}$ versus temperature 
  are presented in Fig.~\ref{Fig3} for $f_{\text{VO}} = 0.2, 0.5$, and $0.8$, for both heating and cooling phases. 
  The real part of $\eps_{\text{mix}}$ is positive valued across the entire temperature range for all volume fractions considered.
  When $f_{\text{VO}} = 0.2$, $\mbox{Im} \lec \eps_{\text{mix}} \ric < 0$ across the entire temperature range; therefore, the homogenized mixture is effectively an active dielectric material for $f_{\text{VO}} = 0.2$.
  When $f_{\text{VO}} = 0.8$, $\mbox{Im} \lec \eps_{\text{mix}} \ric > 0$ across the entire temperature range; therefore, the homogenized mixture is effectively a dissipative dielectric material for $f_{\text{VO}} = 0.8$.
   When $f_{\text{VO}} = 0.5$, $\mbox{Im} \lec \eps_{\text{mix}} \ric < 0$ at low temperatures (less than $63\degC$ for the heating phase and
   less than $32\degC$ for the cooling phase), and
    $\mbox{Im} \lec \eps_{\text{mix}} \ric > 0$ at high temperatures. Therefore, 
    for $f_{\text{VO}} = 0.5$,
    the homogenized mixture is effectively an active material at low temperatures and effectively a dissipative material at high temperatures.
  
  The relative permittivity of the plasmonic material that occupies the half-space $z<0$, namely silver, was taken to be $\eps_{\text{Ag}} = -23.40 + 0.39 i$. Note that $\eps_{\text{Ag}}$ at $\lambdao = 710$ nm is sufficiently insensitive to temperature over the range $25 \degC < T < 80 \degC$  that its temperature dependence need not be considered here \c{Ferrera}.
  
  \section{Surface-plasmon-polariton waves}
  
\begin{figure}[!htb]
\centering
\includegraphics[width=8cm]{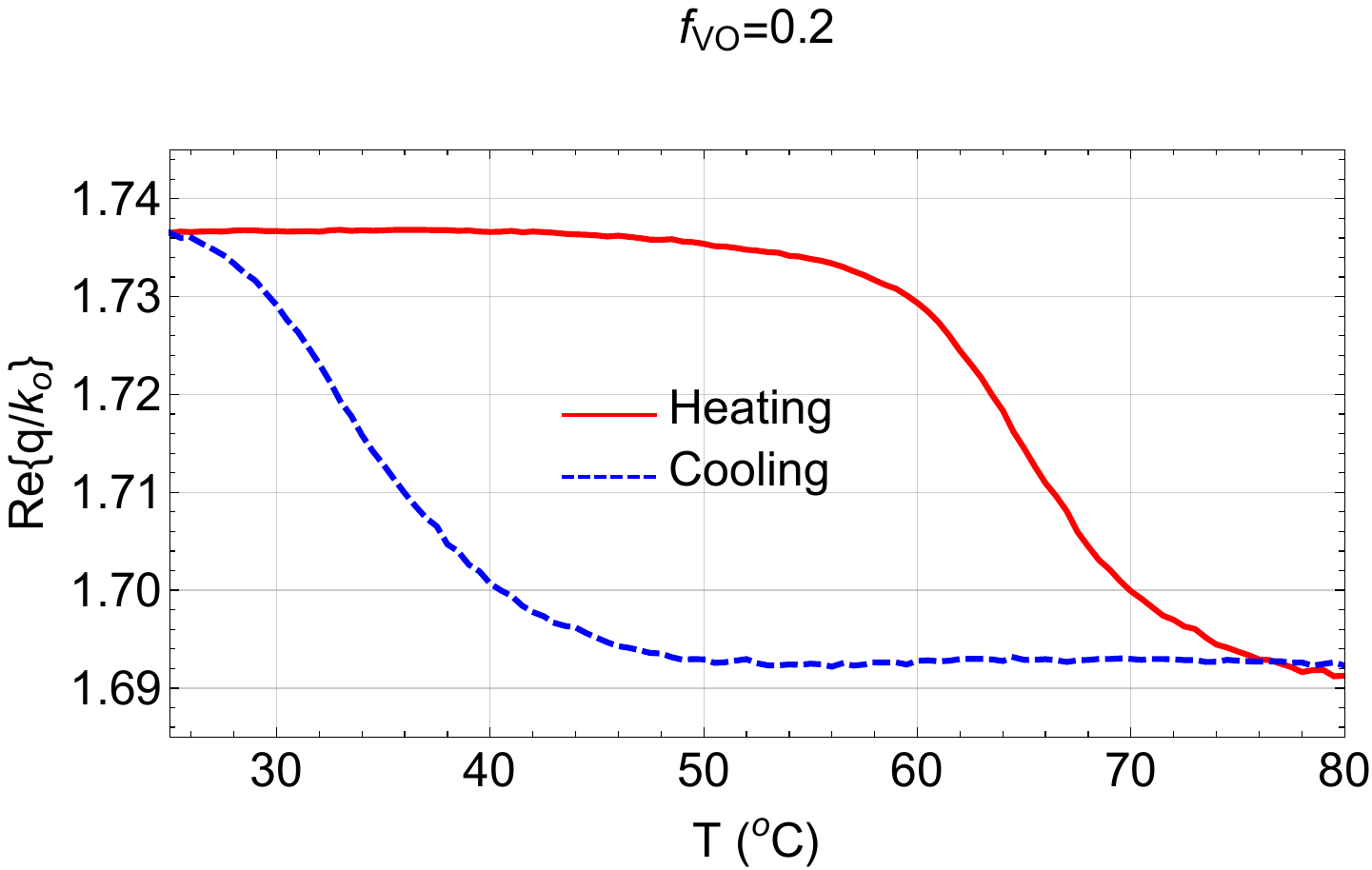}
\hfill
\includegraphics[width=8cm]{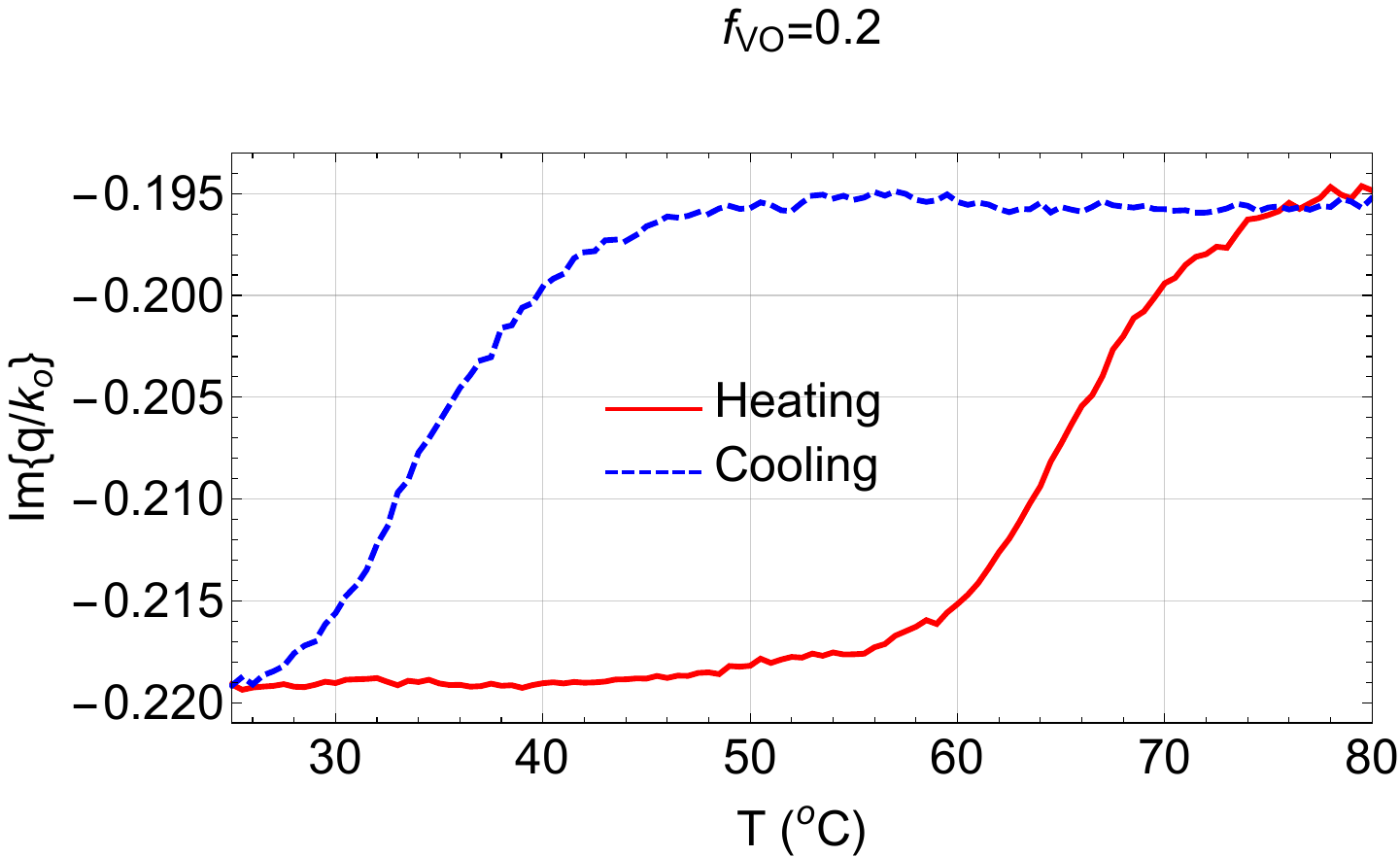}  \vspace{10mm}  \\
\includegraphics[width=8cm]{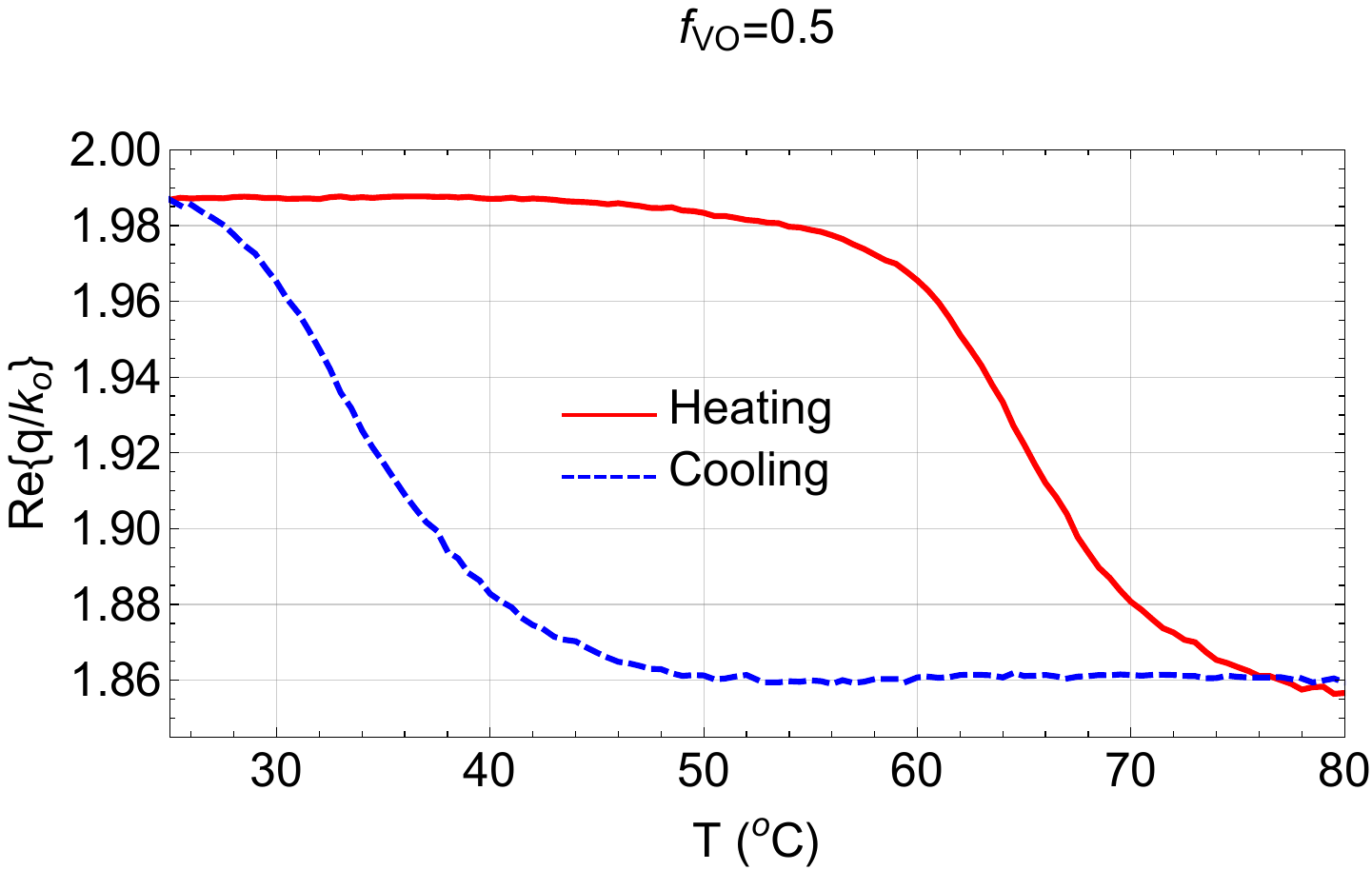}
\hfill
\includegraphics[width=8cm]{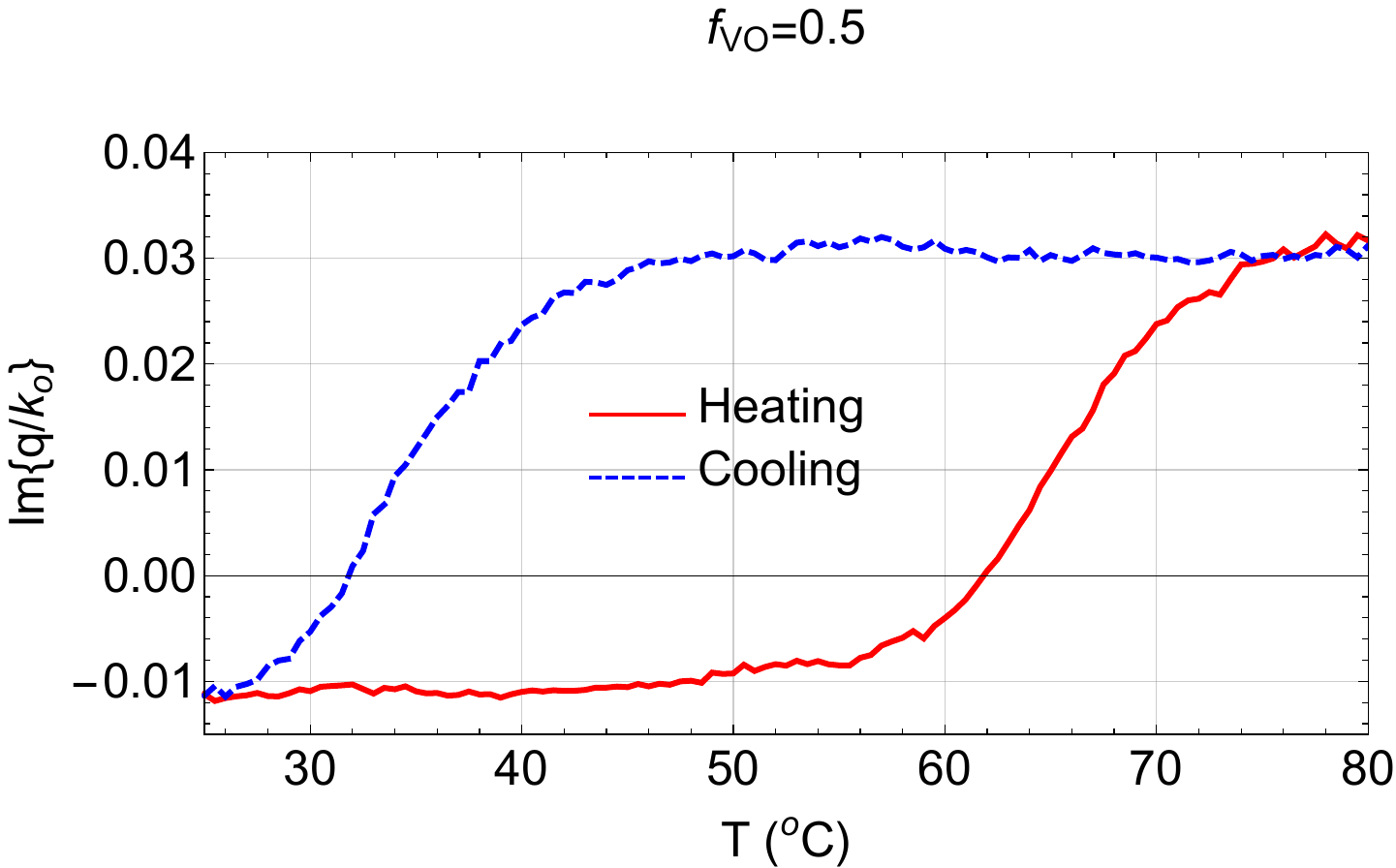} \vspace{10mm} \\
\includegraphics[width=8cm]{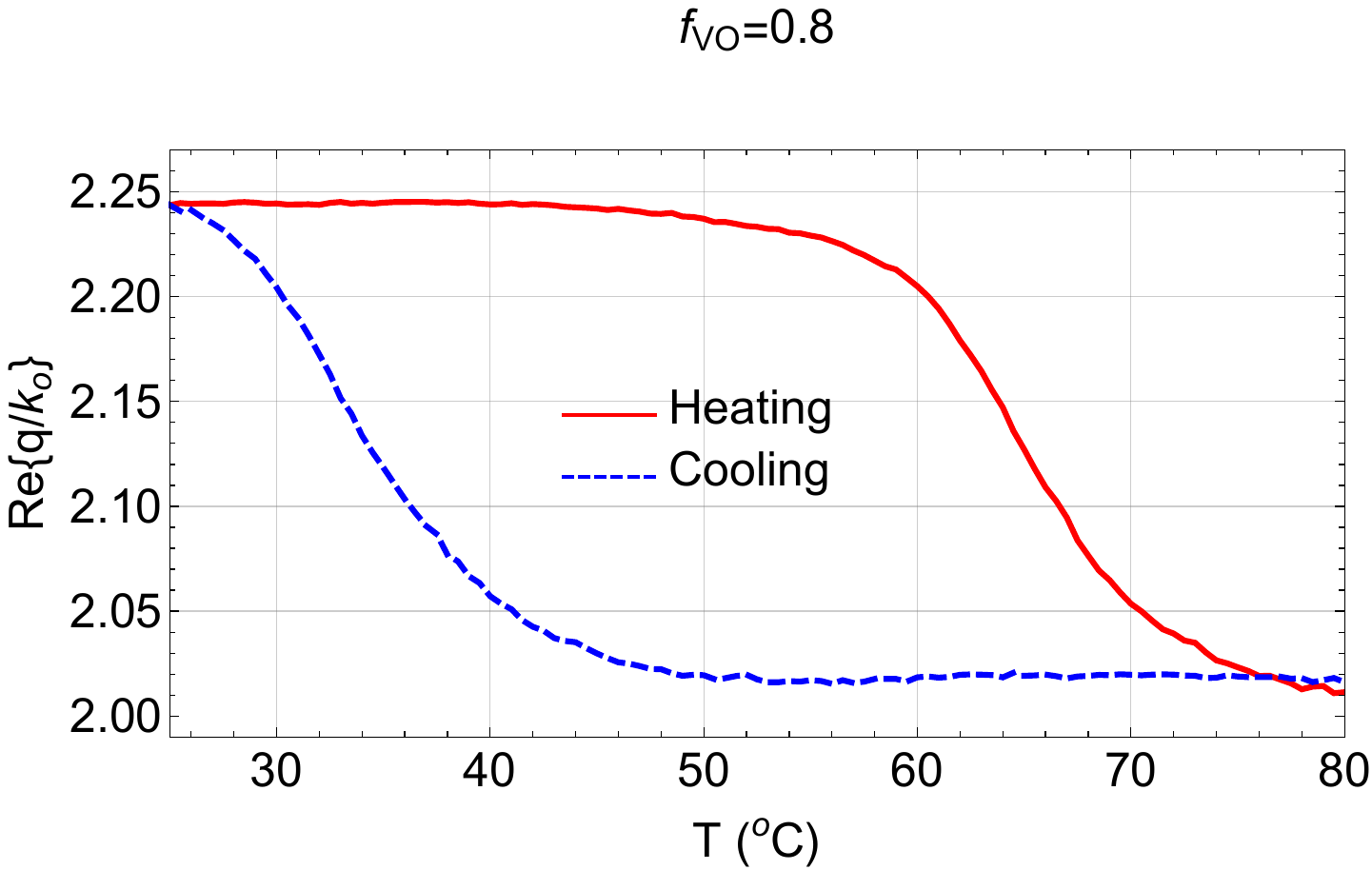}
\hfill
\includegraphics[width=8cm]{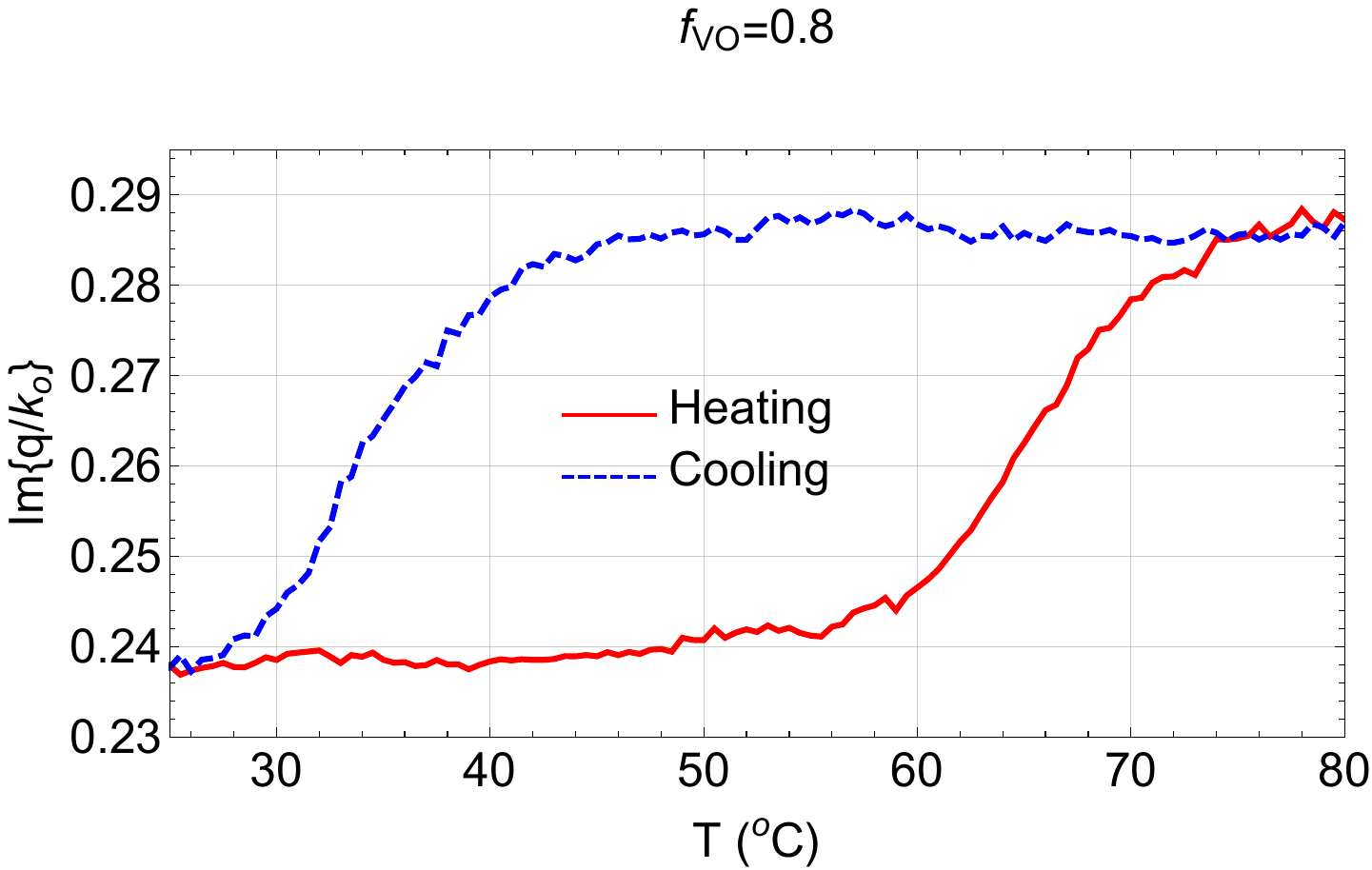}  
     \caption{\label{Fig4} Real and imaginary parts of $q/\ko$ plotted against $T$
     for both heating and cooling phases,
      as determined from Eq.~\r{q} for $f_{\text{VO}}= 0.2$, 0.5, and 0.8.
   }
\end{figure}

  For the canonical boundary-value problem, the wave number of the SPP wave is given by
  \begin{equation} \l{q}
  q = \ko \sqrt{\frac{\eps_{\text{VO}}\eps_{\text{Ag}}}{\eps_{\text{VO}}+ \eps_{\text{Ag}}} },
  \end{equation}
  wherein $\ko = 2 \pi /\lambdao$ is the free-space wave number.
   The real part of $q$ is inversely proportional to the phase speed of the SPP wave, while the
   imaginary part of $q$ is a measure of the SPP wave's attenuation rate, with $\mbox{Im} \lec q \ric <0 $ signifying amplification and $\mbox{Im} \lec q \ric > 0 $ signifying attenuation.
 The real and imaginary parts of $q$ are plotted against temperature for the range $\les 25\degC, 80\degC \ris$  in Fig.~\ref{Fig4} for both heating and cooling phases. The volume fractions considered are
  $f_{\text{VO}} = 0.2, 0.5$, and $0.8$. 
 The real part of $q$ is positive valued across the entire temperature range  for all volume fractions considered. Since, at each temperature, $\mbox{Re} \lec q \ric $ is greater for the heating phase than for the cooling phase, SPP waves propagate at a lower phase speed for the heating phase than for the cooling phase.
  When $f_{\text{VO}} = 0.2$, $\mbox{Im} \lec q \ric < 0$ across the entire temperature range; therefore, the SPP wave is amplified at all temperatures for $f_{\text{VO}} = 0.2$ and the degree of
  amplification is greater if the temperature is increasing rather than decreasing.
  When $f_{\text{VO}} = 0.8$, $\mbox{Im} \lec q \ric > 0$ across the entire temperature range; therefore, the SPP wave is attenuated at all temperatures for $f_{\text{VO}} = 0.8$ and the degree of attenuation is greater if the temperature is decreasing rather than increasing.
   When $f_{\text{VO}} = 0.5$, $\mbox{Im} \lec q \ric < 0$ at low temperatures (less than $63\degC$ for the heating phase and
   less than $32\degC$ for the cooling phase), and
    $\mbox{Im} \lec q \ric > 0$ at high temperatures. Therefore, 
    for $f_{\text{VO}} = 0.5$,
    at a given temperature, whether the SPP wave is amplified or attenuated depends upon whether
    the temperature is increasing or decreasing.
 In particular, at $f_{\text{VO}} = 0.5$, the SPP wave is neither attenuated nor amplified at (i) $T= 63 \degC$ if the temperature is increasing; and (ii) 
 $T= 32 \degC$ if the temperature is decreasing.
 
 \section{Closing remarks}
 
 The propagation of SPP waves at the planar metal/dielectric interface can be controlled by temperature
 by choosing a dielectric material whose constitutive properties are strongly temperature dependent
 and which is impregated with an active dye. Specifically, if the dielectric material is a homogenized mixture of vandium multioxide and rhodamine dyes and the metal is silver, 
 then
  either attenuation or amplification of  the SPP waves may be achieved,
  depending upon the volume fraction of vanadium multioxide.
  The degree of attenuation or amplification is strongly dependent on both the temperature and whether the temperature is increasing or decreasing.
 At intermediate volume fractions of vanadium multioxide, for a fixed temperature, 
  a SPP wave may experience attenuation if the temperature is increasing but experience amplification if the temperature is decreasing.
 This
  thermal hysteresis in amplification and attenuation of SPP waves may be usefully expoited
  in applications involving reconfigurable and multifunctional devices, as well as those involving  temperature sensing.

 \vspace{10mm}
 
 \noindent {\bf Acknowledgments:} 
TGM was supported  by
EPSRC (grant number EP/V046322/1).  AL   was supported by the US National Science Foundation (grant number DMS-1619901)  as well as   the Charles Godfrey Binder Endowment at Penn State.\\


\newpage

\newpage

\newpage

\newpage

\begin{thebibliography}{99}



\bibitem{Boardman}
 Boardman A D (ed) 1982 \emph{Electromagnetic Surface Modes}
(Chicester, Surrey, UK: Wiley)

\bibitem{Pitarke}
Pitarke J M,  Silkin V M,   Chulkov E V,  and  Echenique P M 2007
Theory of surface plasmon and surface-plasmon polaritons \textit{Rep. Prog. Phys.} {\bf 70}  1--87

 \bibitem{ESW_book}
Polo J A  Jr,   Mackay T G,  and  Lakhtakia A 2013 \emph{Electromagnetic Surface Waves: A Modern Perspective} (Waltham, MA, USA: Elsevier)

 \bibitem{Homola}
Homola J (ed) 2006 \emph{ 
 Surface Plasmon Resonance Based Sensors} 
(Berlin, Germany: Springer)


\bibitem{Wang}
  Wang Q,   Rogers E T,  Gholipour B,   Wang C-M,
Yuan G,   Teng J, and   Zheludev N I 2016 Optically reconfigurable metasurfaces and photonic devices based on phase
change materials  \emph{ Nat. Photon. } {\bf 10}   60--65 


\bibitem{Maguid}
Maguid E,   Yulevich I,   Yannai M,   Kleiner V,
 Brongersma M L, and   Hasman E, 2017  Multifunctional interleaved
geometric-phase dielectric metasurfaces \emph{ Light Sci. Appl.}
{\bf  6}  e17027 



\bibitem{Huang}  Huang W,  Yin X,  Huang C,  Wang Q,  Miao T, and  Zhu Y 2010  Optical switching of a
metamaterial by temperature controlling  \emph{ Appl. Phys. Lett.} {\bf 96}  261908 

\bibitem{Waleed}
Waseer W I and Lakhtakia A 2022 Thermal-hysteresis-affected surface-plasmon-polariton-wave propagation
\emph{Mater. Lett.} {\bf 324} 132648

\bibitem{Seo}
  Seo G,  Kim B,  Lee Y W,  and  Kim H 2012  Photo-assisted bistable switching using Mott
transition in two-terminal VO${}_2$ device \emph{ Appl. Phys. Lett.} {\bf 100}  011908  

\bibitem{Cueff}  Cueff S,   John J,  Zhang Z,  Parra J,  Sun J,
Orobtchouk R,  Ramanathan S, and  Sanchis P 2020 VO${}_2$
nanophotonics  \emph{APL Photon.} {\bf 5} 110901  

\bibitem{Lu}
 Lu H,  Clark S,   Guo Y, and   Robertson J 2021
The metal-insulator phase change in vanadium dioxide and its applications \emph{ J. Appl. Phys.} {\bf 129}  240902  

\bibitem{Shi}  Shi R,   Shen N,  Wang J,  Wang W,   Amini A,   Wang N,
and   Cheng C 2019 Recent advances in fabrication strategies,
phase transition modulation, and advanced applications
of vanadium dioxide \emph{ Appl. Phys. Rev.} {\bf  6} 011312  


\bibitem{Cormier}
Cormier P,
  Son T V,
 Thibodeau J,
Doucet A,
 Truong V-V, and 
  Hach\'e A 2017 
Vanadium dioxide as a material to control light polarization in the
visible and near infrared \emph{Opt. Commun.} {\bf 382} 80--85  


\bibitem{Morin}
  Morin F J 1959  Oxides which show a metal-to-insulator transition at the Neel (\emph{sic})
temperature \emph{Phys. Rev. Lett.}  {\bf 3 } 34--36  

\bibitem{Kepic}
 {Kepi\v{c} P,   Ligmajer F,   Hrto\v{n} M,   Ren H,  de~S.~Menezes L,  Maier S A, and   \v{S}ikola T 2021
Optically tunable Mie resonance VO$_2$~nanoantennas for metasurfaces in the visible
\emph{ACS Photonics} {\bf 8}, 1048--1057  }


\bibitem{HW}
 Hodgkinson I J  and   Wu Q h 1997 \emph{Birefringent Thin Films and
Polarizing Elements}    (Singapore: World Scientific)

\bibitem{STF_Book}
Lakhtakia A and    Messier R 2005  \emph{Sculptured Thin Films:
Nanoengineered Morphology and Optics} (Bellingham, WA, USA: SPIE Press)

\bibitem{Sun}
 Sun L,  Yang X, and  Gao J 2013
 Loss-compensated broadband epsilon-near-zero metamaterials with gain media
\emph{Appl. Phys. Lett.} {\bf  103} 201109 

\bibitem{Campione}
 Campione S,  Albani M, and  Capolino F 2011
Complex modes and near-zero permittivity in 3D arrays of plasmonic nanoshells: loss compensation using gain
\emph{Opt. Mater. Exp.} {\bf 1}  1077-1089 

 
\bibitem{Seidel}
{Seidel J, Grafstr\"{o}m S, and Eng L 2005
Stimulated emission of surface plasmons at the interface between a silver film and an optically pumped dye solution
\emph{Phys. Rev. Lett.} {\bf 94} 177401}

 
\bibitem{Torma}
{T\"{o}rm\"{a} P and  Barnes W L  2015 Strong coupling between surface plasmon polaritons and emitters: a review 
\emph{Rep. Prog. Phys.} {\bf 78} 013901
}


\bibitem{Liu}
{Liu S-Y, Li J,  Zhou F,  Gan L, and  Li Z-Y 2011
Efficient surface plasmon amplification from gain-assisted gold nanorods
\emph{Opt. Lett.} {\bf 36} 1296--1298 
}


\bibitem{Berini}
{Berini P and De Leon I 2012
Surface plasmon-polariton amplifiers and lasers
\emph{Nat. Photon.} {\bf 6} 16--24}



\bibitem{Tran}
 Son T V,  Bulmer K, 
  Hach\'e A, and Bisson J-F 2023 
Absence of hysteresis in $n$-$k$ space during the phase transition of vanadium dioxide
\emph{Opt. Commun.}
{\bf 530} 129130


\bibitem{Ward}  Ward L 2000 \emph{The Optical Constants
of Bulk Materials and Films, 2nd Edition} (Bristol, UK: IOP Publishing)

\bibitem{MAEH}
    Mackay T G and   Lakhtakia A 2020
 \emph{Modern  Analytical Electromagnetic  Homogenization with Mathematica, 2nd Edition} (Bristol, UK: IOP Publishing)

\bibitem{Ferrera}
 Ferrera M,  Magnozzi M,  Bisio F, and  Canepa M 2019 Temperature-dependent permittivity of silver and implications for thermoplastics \emph{Phys. Rev. Mat.} {\bf 3} 105201






\end{thebibliography}
\end{document}